\documentclass[runningheads]{llncs}
\usepackage{tikz,lipsum,lmodern}
\usepackage[most]{tcolorbox}

\usepackage{cite}
\usepackage{amsmath,amssymb,amsfonts}
\usepackage{algorithmic}
\usepackage{graphicx}
\usepackage{textcomp}
\usepackage{color, xcolor, colortbl}
\usepackage[title]{appendix}
\begin{document}

\title{Quantum Algorithm Cards: Streamlining the development of hybrid classical-quantum applications}

\titlerunning{Quantum Algorithm Cards}

\author{Vlad Stirbu\inst{1} \orcidID{0000-0001-9462-5922} 
\and
Majid Haghparast\inst{1} \orcidID{0000-0003-3427-5961} 
}
\authorrunning{V. Stirbu and M. Haghparast}
\institute{
University of Jyväskylä, Jyväskylä, Finland\\
\email{vlad.a.stirbu,majid.m.haghparast@jyu.fi}}

\maketitle

\begin{abstract}
The emergence of quantum computing proposes a revolutionary paradigm that can radically transform numerous scientific and industrial application domains. The ability of quantum computers to scale computations implies better performance and efficiency for certain algorithmic tasks than current computers provide. However, to gain benefit from such improvement, quantum computers must be integrated with existing software systems, a process that is not straightforward. In this paper, we investigate challenges that emerge when building larger hybrid classical-quantum computers and introduce the Quantum Algorithm Card (QAC) concept, an approach that could be employed to facilitate the decision making process around quantum technology.

\end{abstract}

\begin{keywords}
Quantum software, software architecture, software development life-cycle,
developer's experience, quantum algorithm cards (QACs).
\end{keywords}

\section{Introduction}

Quantum computers have demonstrated the potential to revolutionize various fields, including cryptography, drug discovery, materials science, and machine learning, by leveraging the principles of quantum mechanics. However, the current generation of quantum computers, known as noisy intermediate-scale quantum (NISQ) computers, suffer from noise and errors, making them challenging to operate. Additionally, the development of quantum algorithms requires specialized knowledge not readily available to the majority of software professionals. These factors pose a significant entry barrier to leveraging the unique capabilities of quantum systems.

For the existing base of business applications, classical computing has already proven its capabilities across a diverse range of solutions. However, some of the computations they must perform can be accelerated with quantum computing, much like GPUs are used today. Therefore, quantum systems should not function in isolation, but they must coexist and inter-operate with classical systems. To this end, software architects play a crucial role in achieving seamless integration while simultaneously designing systems that effectively meet the unique requirements of businesses.

To address the challenges associated with this integration, this paper focuses on designing hybrid systems that integrate quantum and classical computing, aiming to overcome architectural, design, and operational hurdles. In doing so, we look at the software development lifestyle, the technology stack of hybrid classic-quantum systems, and deployment techniques used today. As a concrete contribution, we propose \textit{quantum algorithm cards} as a mechanism to support decision making process related to quantum technology during the development and deployment of hybrid classic-quantum applications.

The rest of the paper is organized as follows. In Section \ref{sec:background}, we provide the necessary background for the paper. In Section 3, we address architectural concerns associated with the development and deployment of  hybrid classic-quantum applications. In Section 4, we introduce the concept of quantum algorithm cards in detail. The discussion and future plans are drawn in Section \ref{sec:conclusions}.

\section{Background}
\label{sec:background}

\begin{figure*}[t]
    \centering
    \includegraphics[width=\textwidth]{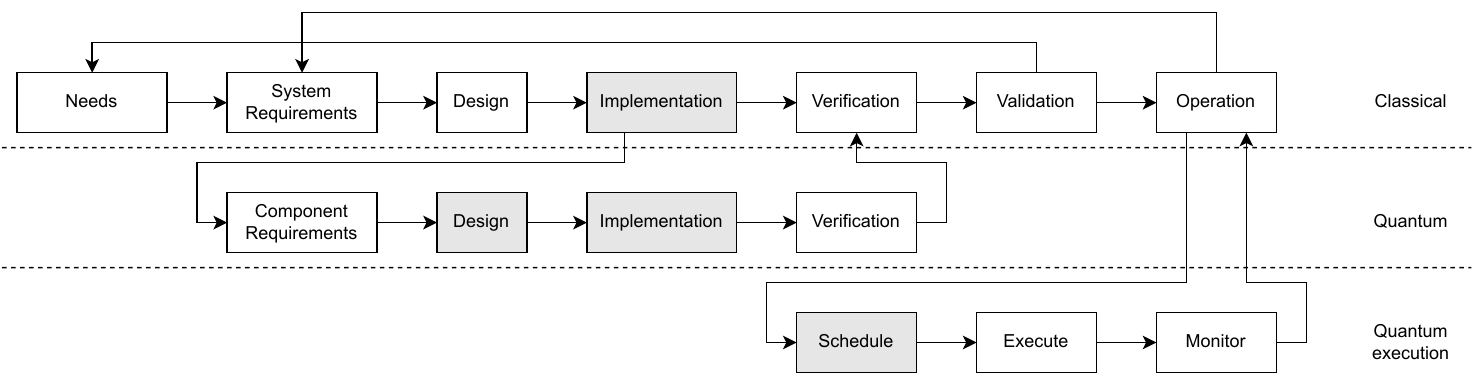}
    \caption{Software development lifecycle of a hybrid classical-quantum system\cite{stirbu2023full}.}
    \label{fig:sdlc}
\end{figure*}

The software development life-cycle (SDLC) of hybrid classic-quantum applications consists of a multi-faceted approach, as depicted in Fig. \ref{fig:sdlc}. At the top level, the classical software development process starts by identifying user needs and deriving them into system requirements. These requirements are transformed into a design and implemented. The result is verified against the requirements and validated against user needs. Once the software system enters the operational phase, any detected anomalies are used to identify potential new system requirements, if necessary. A dedicated track for quantum components is followed within the SDLC \cite{sdlc}, specific to the implementation of quantum technology. The requirements for these components are converted into a design, which is subsequently implemented on classic computers, verified on simulators or real quantum hardware, and integrated into the larger software system. During the operational phase, the quantum software components are executed on real hardware. Scheduling ensures efficient utilization of scarce quantum hardware, while monitoring capabilities enable the detection of anomalies throughout the process.

\begin{figure}[t]
    \centering
    \includegraphics[width=0.6\textwidth]{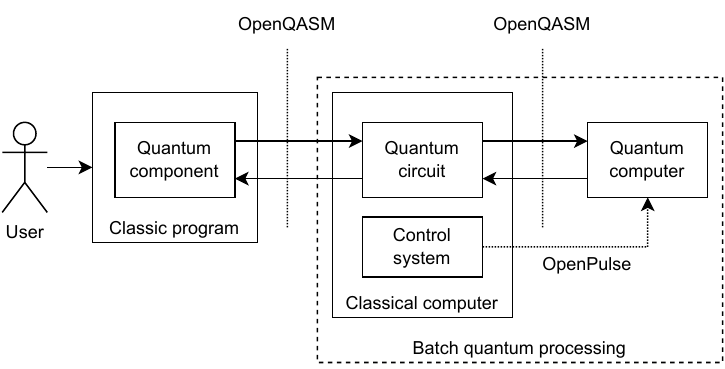}
    \caption{Quantum computing model: components and interfaces}
    \label{fig:system}
\end{figure}

A typical hybrid classic-quantum software system is understood as a classical program that has one or more software components that are implemented using quantum technology, as depicted in Fig. \ref{fig:system}. A quantum component utilizes quantum algorithms \cite{Montanaro2016}, that are transformed into quantum circuits using a toolkit like Cirq\footnote{https://quantumai.google/cirq} or Qiskit\footnote{https://qiskit.org}. The quantum circuit describes quantum computations in a machine-independent language using quantum assembly (QASM) \cite{openqasm}. This circuit is translated by a computer that controls the quantum computer in a machine specific circuit and a sequence of operations, such as pulses \cite{openpulse}, that control the individual hardware qubits. The translation process, implemented using quantum compilers, encompasses supplementary actions like breaking down quantum gates, optimizing quantum circuits, and providing fault-tolerant iterations of the circuit.  
Further, the concept of distributed quantum computers \cite{distributed-quantum-computing}, which interlink multiple distinct quantum machines through quantum communication networks, emerges as a potential solution to amplify the available quantum volume beyond what is possible using a single quantum computer. Nevertheless, the intricacies inherent in the distributed quantum computers remain hidden from users, as compilers aware of the distributed architecture of the target system shield them from such complexities. In essence, the quantum compiler plays a vital role in achieving the effective execution of generic quantum circuits on existing physical hardware platforms, making the compilers an active research area in quantum computing \cite{haghparast2023quantum}.

\section{Architectural and operational concerns}

This section highlights the SDLC stages and the key challenges can be observed while developing hybrid classic-quantum applications.

\textbf{Quantum advantage awareness.} In accordance with the hybrid classic-quantum SDLC, it becomes evident that during the decomposition of the system into smaller software components, a team with limited quantum technology knowledge might overlook that employing quantum algorithms have the capacity to surpass the performance of conventional alternatives for algorithmically intensive tasks.

\textbf{Quantum algorithm design and implementation.} Upon determining the possibility of quantum advantage aligned with their needs, the team often encounters a knowledge gap hindering the effective attainment of this objective. The design of quantum algorithms necessitates comprehension of quantum mechanical phenomena such as superposition, entanglement, and interference – concepts that can confound the intuition of those untrained in the field. Although well-resourced teams will likely have a \textit{quantum scientist} specialty, similar to the data scientist role in artificial intelligence, this cannot be assumed to be generally available. This challenge leaves mainstream developers grappling to identify the optimal algorithms for their tasks. Ultimately, the team resorts to selecting algorithms bundled with widely-used quantum libraries. 

\textbf{Availability and cost of quantum hardware.} Upon successful implementation of the components and confirmation of quantum advantage, the team faces the task of selecting suitable hardware for executing the quantum tasks efficiently. This decision necessitates a comprehensive grasp of the most fitting qubit implementation technology, contingent on the interconnections among the qubits within the generic quantum circuit. While quantum compilers have the capability to convert the initial circuit into a version optimized for the specific machine, the considerations of hardware availability and associated expenses remain pivotal factors to be addressed for each application's context.

\section{Quantum algorithm card proposal}

Addressing the previously highlighted architectural and operational considerations, our proposal recommends incorporating a Quantum Algorithm Card (QAC) in conjunction with quantum algorithm implementations. This artifact purpose is two fold: first, it serves as a repository for insights into the algorithm's implementation, and secondly, it conveys critical information for the users that rely on the implementation to realize the application specific needs. As a starting point, the QAC contains the following sections: the \textit{overview} containing administrative information and high level overview of the algorithm purpose, the \textit{intended purpose} describes the tasks for which the algorithm provide optimal performance, the \textit{usage details} conveys information about how the implementation can be used and integrated into a larger system, the \textit{performance metrics} includes information that is useful to evaluate the results of the algorithm and monitoring for deviations, the \textit{limitations} conveys the known situations for which the use of the algorithm is not suitable, the \textit{ reference} refers to the canonical document that introduced the algorithm, and, finally, the \textit{caveats} should include relevant information that the user should be aware. Table \ref{demo-table} provides an outline of the proposed QAC elements, the corresponding content and the recipients of the information. It's important to note that these elements are not exhaustive and can be tailored to suit the specifics of each corresponding quantum algorithm.

\begin{table}[t]
\caption{\label{demo-table} Quantum algorithm card: elements, content and recipients (e.g. T - technology management and architects, D - software developers, O - operations)}
\centering
\begin{tabular}{ |p{1.8cm}|p{9.2cm}| p{1cm} |}
\hline
\rowcolor{lightgray}
\multicolumn{3}{|c|}{Quantum Algorithm Card} \\
\hline
\rowcolor{lightgray}
\multicolumn{1}{|c|}{Element} & \multicolumn{1}{c|}{Description} & \multicolumn{1}{c|}{Target} \\
\hline
Overview & 
- Provider/designer/maintainer information.

- Brief description of the algorithm's purpose, key features and functionalities.

- Algorithm's high-level architecture or approach, complexity &
T D O
\\

\hline
Intended use &
- Clear description of the tasks the algorithm is designed for.

- Specific scenarios for which the quantum algorithm is intended. &
T
         \\

\hline
Usage \mbox{details}   & 
- Information about algorithm usage, e.g. inputs and outputs.
        
- Quantum volume needed to run the algorithm.
&
D O
\\

\hline
Performance metrics & 
- Metrics and explanations used to evaluate the algorithm's performance.
  
- Decision thresholds, variation approaches, and any relevant quantitative measures. &
T O
\\
\hline
Limitations & 
- Clear articulation of the algorithm's limitations and potential failure modes.
            
- Known scenarios where the algorithm might not perform well or could provide incorrect results. &
T D
\\
\hline
References & 
- Citations to relevant research papers and resources. &
T D
\\
\hline
Caveats & 
- Situations that users should avoid. &
D
\\
\hline
\end{tabular}
\end{table}

\section{Discussion and Future Work}
\label{sec:conclusions}

Serving as a facilitator, the QAC artifact aims to enhance communication across common specializations within the team to support \textit{decision making}. For example, technology managers and architects need specific information during the high-level implementation phase; software engineers need to know how the integrate the quantum technology into the larger classical system; operations needs to know how to execute and monitor the quantum components in production environment. Consequently, it is imperative that the card's content is conveyed in a language that is easily comprehensible by the intended audience, which are users and not developers of quantum technology.

The initial evaluation of the concept was performed, as an internal exercise based on a paper prototype \cite{card-example} of Grover' search algorithm \cite{grover1996fast}, on a target group that has both classic and quantum software development skills. The concept was found to be useful, especially for the developers that have artificial intelligence and machine learning background, as they were already familiar with similar concepts like Model Cards \cite{modelcards}, and Data Cards \cite{datacards}. However, as the classic and quantum disciplines are rather different, there is a fine line that needs to the considered carefully when deciding the depth of content about the quantum algorithm should be included, otherwise the card becomes a communication impediment rather than facilitator.

Further work is needed to validate the concept with external target groups. We are also planning to develop a Python toolkit that aims to streamline the collection of information included, and to automate the generation of QACs. The Quantum Algorithm Cards Toolkit (QACT) will enable the developers and implementers of quantum algorithms share their metadata and metrics with developers, researchers, and other stakeholders.

\section*{Acknowledgement}
This work has been supported by the Academy of Finland (project DEQSE 349945) and Business Finland (project TORQS 8582/31/2022).

\bibliographystyle{ieeetr}
\bibliography{references}

\end{document}